\documentclass{eas}
\usepackage{graphicx}
\usepackage{amssymb}
\begin{document}

\title{Densitometry and Thermometry of Starburst Galaxies}
\runningtitle{Densitometry and Thermometry of Starburst Galaxies}
\author{Jeffrey G.~Mangum}
\address{National Radio Astronomy Observatory, 520 Edgemont Road,
  Charlottesville, VA 22903-2475, USA; \email{\texttt{jmangum@nrao.edu}}}
\author{Jeremy Darling}
\address{Center for Astrophysics and Space Astronomy, Department of
  Astrophysical and Planetary Sciences, Box 389, University of
  Colorado, Boulder, CO 80309-0389, USA;
  \email{\texttt{jdarling@origins.colorado.edu}}}
\author{Karl M.~Menten}
\address{Max Planck Instit\"ut f\"ur Radioastronomie, Auf dem H\"ugel
  69, 53121 Bonn, Germany;
  \email{\texttt{kmenten@mpifr-bonn.mpg.de},\texttt{chenkel@mpifr-bonn.mpg.de}}} 
\author{Christian Henkel}
\sameaddress{3}
\author{Meredith MacGregor}
\sameaddress{1}
\secondaddress{Harvard University, Cambridge, MA 02138, USA; \email{\texttt{mmacgreg@fas.harvard.edu}}}
\begin{abstract}
With a goal toward deriving the physical conditions in external
galaxies, we present a survey of formaldehyde (H$_2$CO) and
ammonia (NH$_3$) emission and absorption in a sample of starburst
galaxies using the Green Bank Telescope.  By extending well-established
techniques used to derive the spatial density in star formation
regions in our own Galaxy, we show how the relative intensity of the
$1_{10}-1_{11}$ and $2_{11}-2_{12}$ K-doublet transitions of H$_2$CO
can provide an accurate densitometer for the active star formation
environments found in starburst galaxies (\cf\ Mangum \etal\ \cite{Mangum2008}).
Similarly, we employ the well-established 
technique of using the relative intensities of the (1,1), (2,2), and
(4,4) transitions of NH$_3$ to derive the kinetic temperature in
starburst galaxies.  Our measurements of the kinetic temperature
constrained spatial density in our starburst galaxy sample represent
the first mean density \textit{measurements} made toward starburst
galaxies.  We note a disparity between kinetic temperature measurements
derived assuming direct coupling to dust and those derived from our
NH$_3$ measurements which points to the absolute need for direct gas
kinetic temperature measurements using an appropriate molecular probe.
Finally, our spatial density measurements point to a rough constancy
to the spatial density (10$^{4.5}$ to 10$^{5.5}$ cm$^{-3}$) in our
starburst galaxy sample.  This implies that the Schmidt-Kennicutt
relation between L$_{IR}$ and M$_{dense}$: (1) Is a measure of the
dense gas mass reservoir available to form stars, and (2) Is not
directly dependent upon a higher average density driving the star
formation process in the most luminous starburst galaxies.

\end{abstract}
\maketitle
\section{Introduction}

Studies of the distribution of carbon monoxide (CO) emission in
external galaxies (\cf\ Young \& Scoville \cite{Young1991}) have
pointed to the presence 
of large quantities of molecular material in these objects.  These
studies have yielded a detailed picture of the molecular mass in many
external galaxies.  But, because emission from the abundant CO
molecule is generally dominated by radiative transfer effects, such as
high optical depth, it is not a reliable monitor of the physical
conditions, such as spatial density and kinetic temperature,
quantities necessary to assess the potential for star formation.
Emission from less-abundant, higher-dipole moment molecules 
are better-suited to the task of deriving the spatial density and
kinetic temperature of the dense gas in our and external galaxies.  For this
reason, emission line studies from a variety of molecules have been
made toward mainly nearby galaxies (see Henkel
\etal\ \cite{Henkel1991} for a review).

Formaldehyde (H$_2$CO) has proven to be a reliable density 
and kinetic temperature probe in Galactic molecular clouds.  
Existing measurements of the H$_2$CO $1_{10}-1_{11}$ and
$2_{11}-2_{12}$ emission in a variety of galaxies by
Baan \etal\ (\cite{Baan1986}, \cite{Baan1990}, \cite{Baan1993}) and
Araya \etal\ (\cite{Araya2004}) have mainly concentrated on measurements of the 
$1_{10}-1_{11}$ transition.  One of the goals of the present study
was to obtain a uniform set of measurements of both K-doublet
transitions with which the physical conditions, specifically the
spatial density, in the extragalactic context could be derived.  Using
the unique density selectivity of the K-doublet transitions of H$_2$CO
within the framework of a radiative transfer model which incorporates
the Large Velocity Gradient (LVG) approximation, we have measured the
spatial density in a sample of galaxies exhibiting starburst phenomena
and/or high infrared luminosity.  Our ammonia (NH$_3$) measurements of
these same galaxies, together with our LVG model, provide kinetic
temperature measurements which can further constrain the range of
allowed densities derived from our H$_2$CO measurements.

\section{Galaxy Sample Summary}

\begin{figure}
\centering
\includegraphics[scale=0.18]{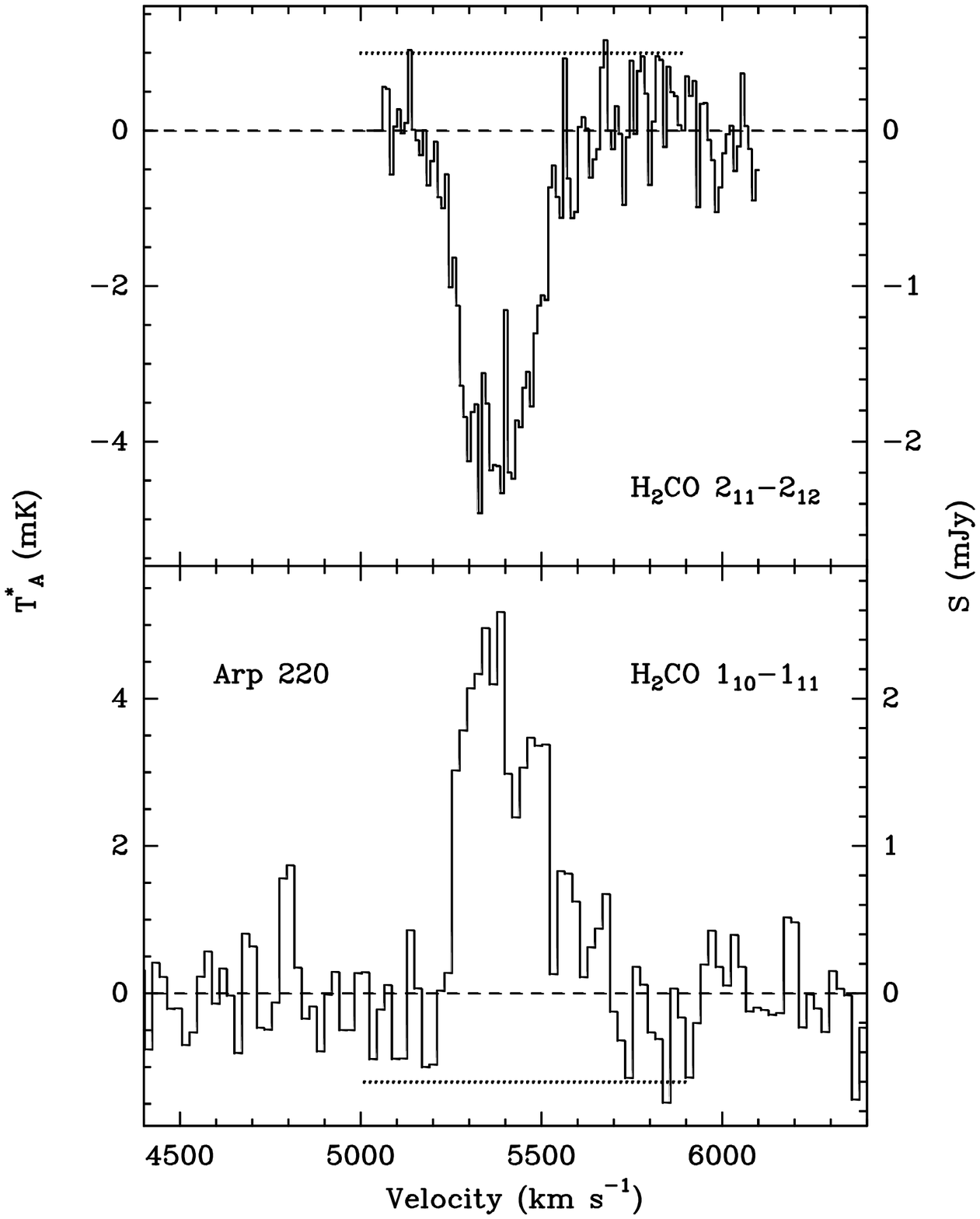}
\includegraphics[scale=0.18]{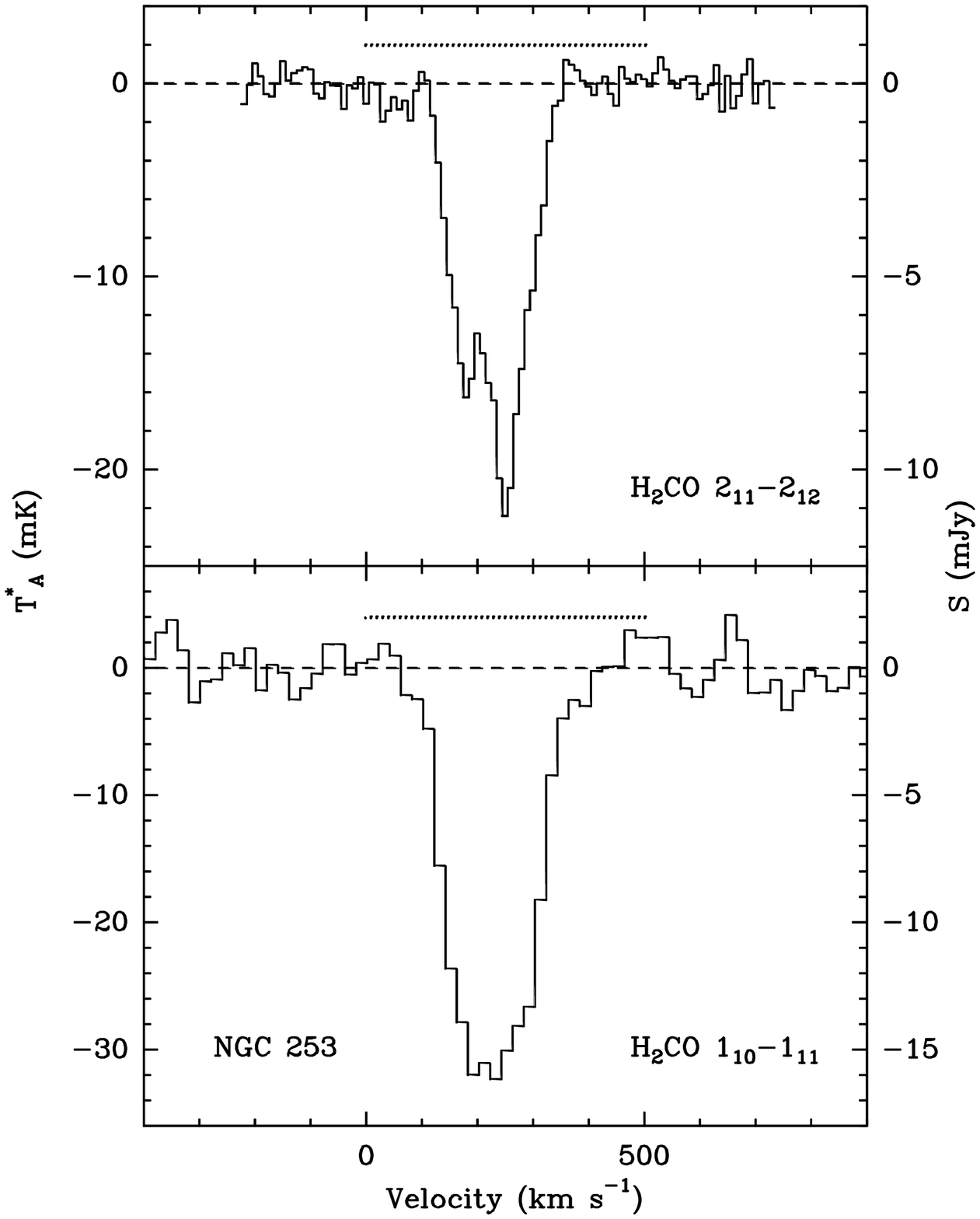} \\
\includegraphics[scale=0.16,viewport=0 20 550 800]{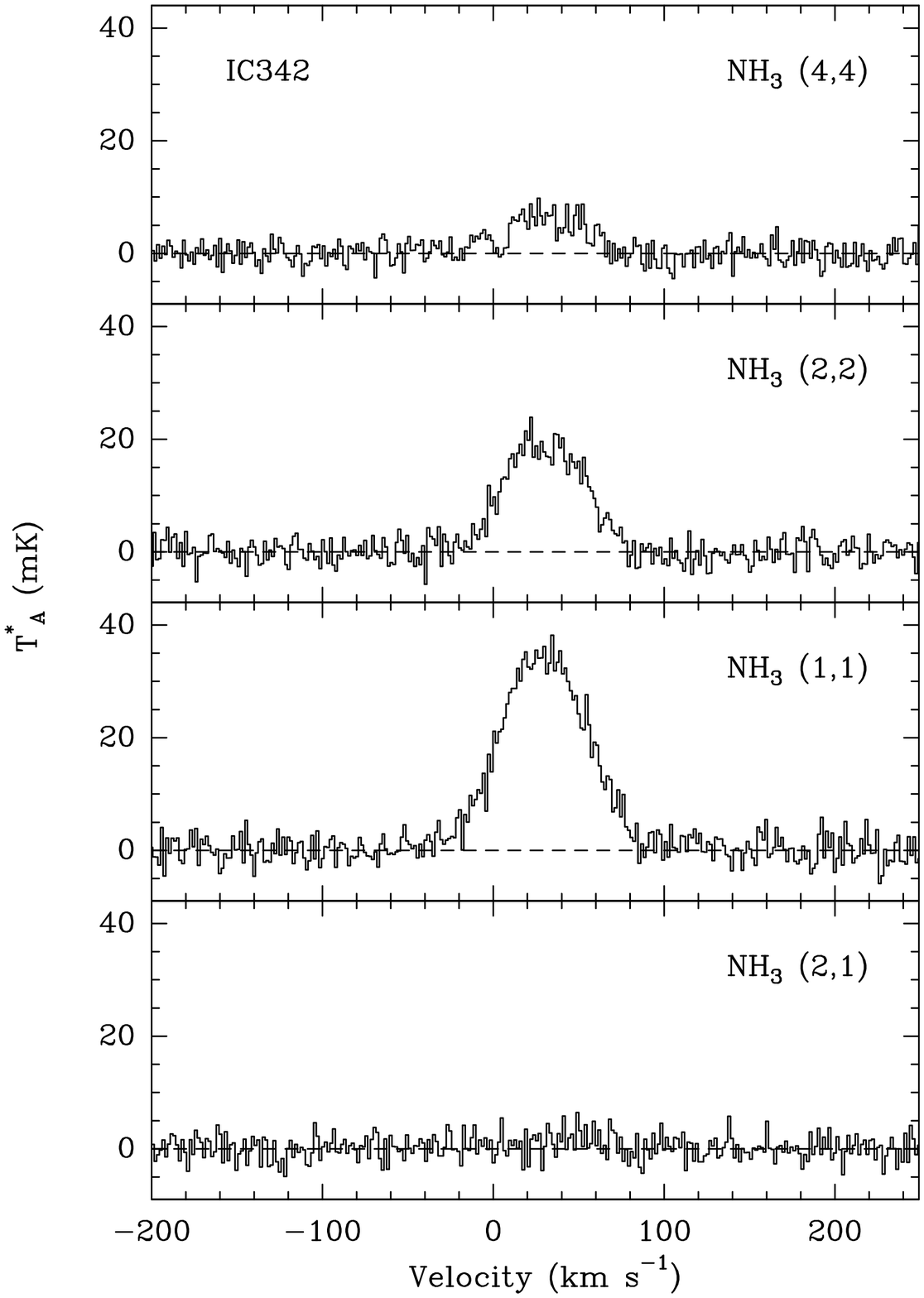}
\includegraphics[scale=0.16,viewport=0 20 550 800]{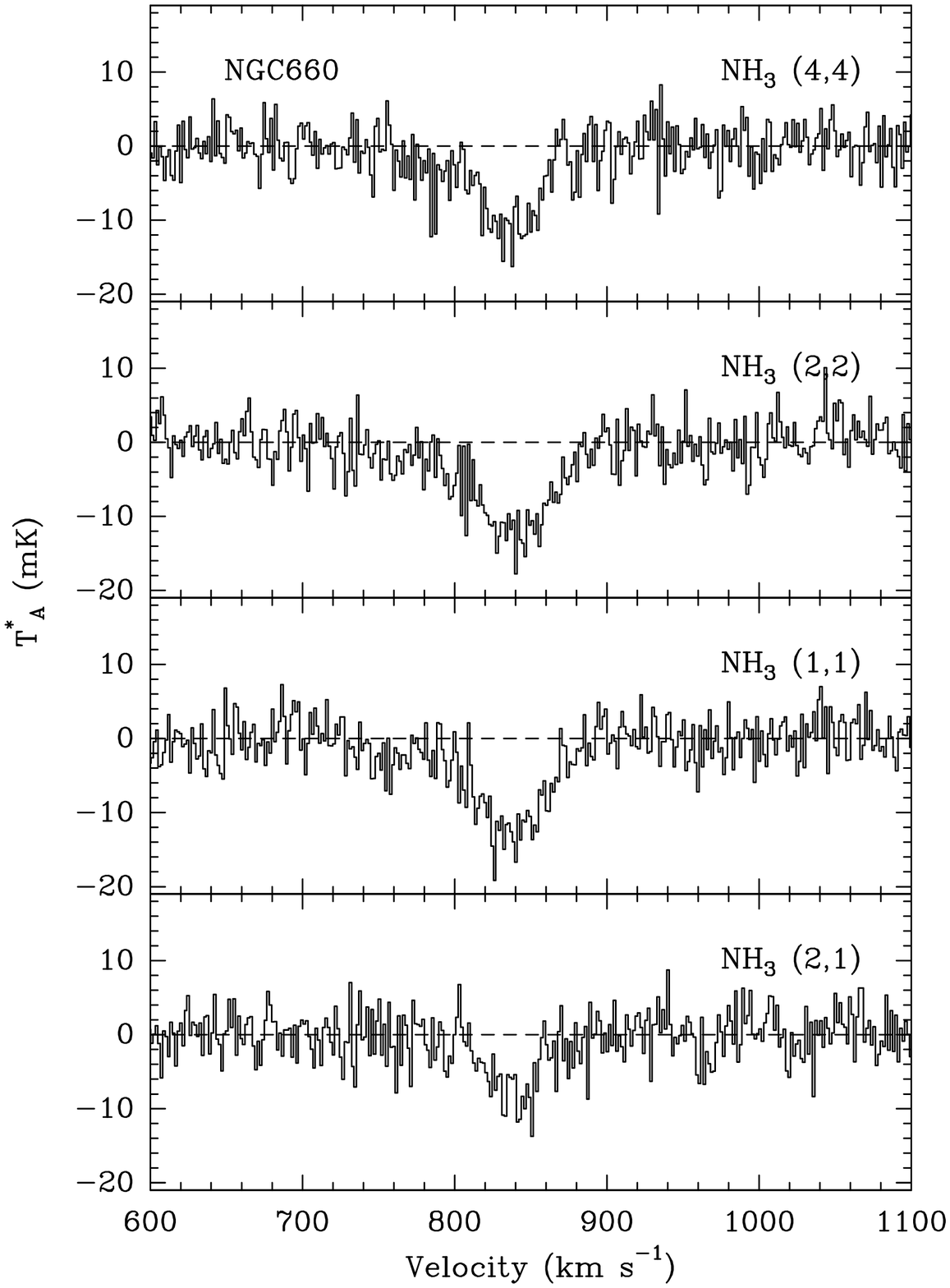}
\caption{Top Row: H$_2$CO $1_{10}-1_{11}$ (top) and
  $2_{11}-2_{12}$ (bottom) spectra from Arp~220 (left) and NGC~253
  (right).  Bottom Row: NH$_3$ (4,4), (2,2), (1,1), and (2,1) 
  (top-to-bottom) spectra from IC~342 (left) and NGC~660 (right).}
\label{fig:ExgalSamples}
\end{figure}

The galaxy selection criteria and summary measurement results are as
follows: Selection Criteria: Dec(J2000) $> -40^\circ$; L(60 $\mu$m)
$>$ 50 Jy and/or bright HCN and CO emission.  H$_2$CO measurements
summary: GBT $\theta_b$ = 51 arcsec (\@$\lambda$ = 2 cm) and 153
arcsec (\@$\lambda$ = 6 cm); 26 of 56 galaxies detected in at least
one H$_2$CO transition; $1_{10}-1_{11}$ emission/absorption = 6/19;
$2_{11}-2_{12}$ emission/absorption = 1/13; 17 of 26 detections are
new extragalactic discoveries of H$_2$CO.  Figure~\ref{fig:ExgalSamples}
shows two examples of our H$_2$CO results.  Emission in
either of the two H$_2$CO transitions in this study is a sign of high
spatial density.  Observed emission in the H$_2$CO $2_{11}-2_{12}$
transition toward Arp~220 signals a relatively high density
($\log(n(H_2)(cm^{-3})) \gtrsim 5.0$).  Absorption in both H$_2$CO
transitions toward NGC~253, on the other hand, signals an average
density that is lower ($\log(n(H_2)(cm^{-3})) \lesssim 5.0$).

NH$_3$ measurements summary: GBT $\theta_b$ = 30 ($\lambda$ = 1.3 cm)
arcsec; 12 of 19 galaxies detected in at least one NH$_3$ transition;
NH$_3$ emission/absorption = 7/5; 4 of 12 detections are new
extragalactic discoveries of NH$_3$.  Figure~\ref{fig:ExgalSamples}
shows two examples of our NH$_3$ measurement results.  As these
examples reveal, our NH$_3$ measurements reveal both emission and
absorption in the (1,1), (2,2), (4,4), and absorption in the (2,1)
transition.  \textit{Emission} in the NH$_3$ (2,1) transition is a
hallmark of high infrared intensity (\cf\ Morris
\etal\ \cite{Morris1973}).  NH$_3$ (2,1) absorption, on the other
hand, requires only a strong background continuum source.

\section{Density and Kinetic Temperature Measurements}

For the H$_2$CO sources in which we have detected both the
$1_{10}-1_{11}$ and $2_{11}-2_{12}$ transitions we have applied our
technique to derive the spatial density assuming kinetic temperatures
derived both from dust emission and our NH$_3$ measurements
(Table~\ref{tab:dentemps}).  Noteworthy are the generally higher, and
in a few cases \textit{much} higher, kinetic temperatures derived
using our NH$_3$ measurements.  In many cases these high kinetic
temperatures have been previously noted (\cf\ Ott
\etal\ \cite{Ott2005}), pointing to the importance of a direct
measurement of the kinetic temperature in these starburst environments.

\section{Conclusions}

The H$_2$CO and NH$_3$ measurements presented allow us to derive the
kinetic temperature and 
spatial density in our starburst galaxy sample, representing the first
mean density \textit{measurements} made toward starburst galaxies.
The disparity between kinetic temperature measurements derived assuming
direct coupling to dust and those derived from our NH$_3$ measurements
points to the absolute need for direct gas kinetic temperature
measurements using an appropriate molecular probe.  Finally, our
spatial density measurements point to a rough constancy to the spatial
density (10$^{4.5}$ to 10$^{5.5}$ cm$^{-3}$) in our starburst galaxy
sample.  This implies that the Schmidt-Kennicutt relation between
L$_{IR}$ and M$_{dense}$: (1) Is a measure of the dense gas mass
reservoir available to form stars, and (2) Is not directly dependent
upon a higher average density driving the star formation process in
the most luminous starburst galaxies.

%
\begin{table}
\scriptsize
\centering
\caption{Derived Kinetic Temperatures and Spatial Densities}
\begin{tabular}{|lccclc|}
\hline
Galaxy & $\alpha$(J2000) & $\delta$(J2000) & v$_{hel}$ &
T$_D$,T$_K$ & log(n(H$_2$)) \\
&&& (km s$^{-1}$) & (K) & (cm$^{-3}$) \\
\hline
NGC~253       & 00:47:33.1  & $-$25:17:18 & 251 & 34 (D) &
$5.08\pm0.02$ \\
              &&&& $81\pm2$ & $5.00\pm0.03$ \\
IC~342        & 03:46:49.7  & $+$68:05:45 & 31  & 30 (D)
& $5.05\pm0.11$ \\
              &&&& $38\pm3$ & $5.05\pm0.11$ \\
              &&&& $104\pm7$ & $4.89\pm0.14$ \\
M~82          & 09:55:52.2  & $+$69:40:47 & 203 & 45 (D) &
$4.95\pm0.02$ \\  
              &&&& $58\pm16$ & $4.92\pm0.05$ \\
NGC~3079C1    & 10:01:57.8  & $+$55:40:47 & 1150 & 32 (D) &
$5.56\pm0.02$ \\  
              &&&& $>260$ & $<4.42$ \\
NGC~3079C2    &&&& 32 & $5.47\pm0.02$ \\
              &&&& $>180$ & $<4.94$ \\
IC~860        & 13:15:04.1  & $+$24:37:01 & 3866   & 40 (D) &
$5.70\pm0.02$ \\
              &&&& $>52$ & $<5.67$ \\
M~83          & 13:37:00.9  & $-$29:51:57 & 518  & 31 (D) &
$5.38\pm0.08$ \\  
              &&&& $48\pm9$ & $5.39\pm0.09$ \\
Arp~220       & 15:34:57.1  & $+$23:30:11 & 5434 & 44 (D) &
$5.64\pm0.02$ \\  
              &&&& $>180$ & $<5.00$ \\
NGC~6946      & 20:34:52.3  & $+$60:09:14 & 48   & 30 (D) &
$5.05\pm0.20$ \\
              &&&& $44\pm7$ & $5.06\pm0.21$ \\
\hline
\multicolumn{6}{l}{D indicates dust temperature (see Mangum
  \etal\ \cite{Mangum2008}).}
\end{tabular}
\label{tab:dentemps}
\end{table}

\end{document}